\documentclass[reprint,prl,superscriptaddress]{revtex4-1}
\usepackage{color}
\usepackage{array}
\usepackage{amsthm}
\usepackage{amsmath}
\usepackage{amssymb}  
\usepackage{graphicx}
\usepackage{esint}

\usepackage{H1}
\begin{document}
\newcommand{\dlsq}{\left[\!\left[}
\newcommand{\drsq}{\right]\!\right]}
\renewcommand{\arraystretch}{1.5}

\title{Enhanced bulk-edge Coulomb coupling in Fractional Fabry-Perot interferometers}
\author{C.W.~von~Keyserlingk}
\email{curtvk@princeton.edu}
\affiliation{Princeton Center for Theoretical Science, Princeton University, Princeton, New Jersey 08544, USA}
\affiliation{Rudolf Peierls Centre for Theoretical Physics, 1 Keble Road, Oxford, OX1 3NP, United Kingdom}
\author{S.H.~Simon}
\email{s.simon1@physics.ox.ac.uk}
\affiliation{Rudolf Peierls Centre for Theoretical Physics, 1 Keble Road, Oxford, OX1 3NP, United Kingdom}
\author{Bernd Rosenow}
\email{rosenow@physik.uni-leipzig.de}
\affiliation{Institut f\"ur Theoretische Physik, Universit\"at Leipzig, D-04103, Leipzig, Germany}
\date{\today}
\begin{abstract}
We study the effects of bulk-edge Coulomb coupling on quantum Hall Fabry-Perot interferometers. We find that these effects can be appreciable in devices which would not usually be associated with strong bulk-edge Coulomb coupling, provided the devices in question exhibit certain fractional plateaus. With this in mind, we analyze recent experiments at $\nu=5/2$ by taking into account a  tunnel coupling between localized bulk Majorana states and Majorana edge states. We find that these experimental data are consistent with the widely held view that the $\nu=5/2$ state harbors Moore-Read topological order. However, experiments may have measured Coulomb effects rather than an `even-odd effect' due to non-abelian braiding.
\end{abstract}

\maketitle
The Fabry-Perot (FP) interferometer has proved to be a useful tool for probing the physics of both integer and fractional quantum Hall states \cite{Chamon97}. A FP interferometer  is a Hall bar perturbed by two constrictions, which introduce amplitudes for interedge tunneling of quasiparticles (QPs) and thus give rise to interference. The probability of backscattering is a function of magnetic field and the interferometer's area.
For idealized devices the interference measurements contain information about the (potentially fractional) charges of interfering QPs, and their braiding with localized quasiparticles in the bulk. However, the real devices are more complicated, in part due to the Coulomb coupling (CC) between the bulk and the edge \cite{Rosenow2011,Rosenow2007}. Two regimes were identified which had rather different interference traces: Aharonov-Bohm (AB) where bulk and edge are decoupled, and `Coulomb dominated' (CD) where they are very strongly coupled \cite{Zhang2009,Ofek2010,Choi2011}. In this work we highlight the fact that CD-like behavior can be present in a device with relatively weak bulk-edge CC, provided the device exhibits certain fractional plateaus.  

FP interferometry has been used to test the theoretical prediction that the experimentally observed $\nu=5/2$ Fractional Quantum Hall Effect (FQHE) plateau is described by a Moore-Read (MR) Pfaffian (PF) or anti-Pfaffian ($\overline{\text{PF}}$) state suggested by theory \cite{Moore91,Lee07,Levin07}. Both of these candidate states support non-abelian QP excitations. While this putative MR state cannot be used for universal quantum computation \cite{Nayak08}, it could well be the first experimentally realized non-abelian topological phase, and verifying this is an important proof of concept step on the road to topological quantum computation. If non-abelian topological order is present, then the simplest models for these devices predicts the longitudinal resistance of the interferometer should oscillate in a manner that depends sensitively on the parity of the number of QP's in the FP cell (the `even-odd' effect) \cite{Stern06,Bonderson06}. Experimental groups have attempted to confirm the presence of these oscillations \cite{Willett1,Willett2,Willett3,Willett4,Willett5,Kang11}, although the interpretation of these experiments are still in dispute. In part of this work, we discuss the extent to which bulk-edge Coulomb coupling is compatible with these experimental results. 

This work is organized as follows. We briefly overview the effect of bulk-edge CC, and its experimental signatures for a variety of Abelian integer and fractional quantum Hall states, highlighting in particular how CC can be augmented in high mobility samples, where many fractional plateaus are present. Then we discuss the $\nu=5/2$ state, emphasizing the consequences of a tunnel coupling between bulk and edge Majorana modes. 
We then incorporate moderate bulk-edge CC into these theoretical models, and find that the results compare favorably to recent experiments \cite{Willett1,Willett2,Willett3,Willett4,Willett5}, although our model conflicts strongly with previous interpretations given to the data.

\paragraph*{\bf Bulk-edge Coulomb coupling and abelian plateaus\label{sec:Coulomb-effects}}

-- In the standard picture of the FP interferometer, the area enclosed by the interfering edge is assumed to vary little and smoothly upon adding a few flux quanta to the cell. However, this picture ignores the CC between
the charges in the plateau and in the compressible edge \cite{Rosenow2011}. For a device at FQHE filling $\nu_{\text{in}}$, the central bulk has charge 
\be\label{eq:cbc}
e_{\text{in}}^{*}N_{\text{L}}+\nu_{\text{in}}\frac{B A}{\phi_{0}}-\bar{q}
\ee

 where $e^{*}_{\text{in}}$ is the charge of the fundamental QP (in units of electron charge $e<0$), $\nu_{\text{in}}$ is the filling fraction of the central plateau, $BA$ is the total magnetic flux through the central plateau region, $\phi_{0}=h/e$ is the magnetic flux quantum, and $\overline{q}$ is the total effective background charge.  The number of localized QP's in the bulk is (up to an unimportant constant)
 \be\label{eq:NL}
 N_{\text{L}}=\left[\!\left[ \frac{\overline{q}}{e_{\text{in}}^{*}}- \frac{\nu_{\text{in}}B A}{e_{\text{in}}^{*}\phi_{0}}\right]\!\right]\punc{,}
 \ee
 where $\left[\!\left[ x \right]\!\right]$ denotes the integer closest to $x\in\mathbb{R}$. The compressible edge changes size in sympathy to the central bulk charge \eqnref{eq:cbc}. For instance, a small increase in magnetic field leads to an increase in electron density in the cell, which has the effect of depleting charges from the edge, decreasing the area enclosed by the interfering edge QP's. The sudden appearance of a QP in the bulk has the same effect. As a result, the total area of the central plateau varies around an (approximately) constant value $A_{0}$ like $A=A_{0}+\delta A$ . The resulting phase acquired by a fundamental QP encircling the FP cell at zero temperature is \cite{Rosenow2011}
\begin{align}
\frac{\theta}{2\pi}=  e_{\text{in}}^{*}\gf-\frac{\kappa e_{\text{in}}^{*}}{\Delta \nu}(e_{\text{in}}^{*}N_{\text{L}}+\nu_{\text{in}}\phi-\bar{q})+N_{\text{L}}\frac{\theta_{a}}{2\pi}\punc{,}
\label{eq:phase}
\end{align}
where $\phi=BA_{0}$, $\Delta \nu=\nu_{\text{in}}-\nu_{\text{out}}$ is the difference in filling fraction between the central and next outermost plateau, and $\theta_{a}$ is the monodromy due to braiding two fundamental QP's in the $\nu_{\text{in}}$ plateau.  The CC between the bulk and the edge is quantified by $\kappa\geq0$ -- in particular we recover the `Aharonov-Bohm (AB) limit' when CC is absent i.e., when $\kappa=0$. Note that the second term on the right hand side has an inverse dependence on $\Delta \nu$, so that the effect of CC is enhanced when the next outermost plateau is nearby in filling. The longitudinal resistance for the Fabry-Perot interferometer is predicted to vary like $\delta R_{\text{L}} \propto \Re [e^{i \theta}]$. With this in mind, we now summarize how we expect $R_{\text{L}}$ to vary if we change $B$ or $V_{\text{G}}$. It can be shown that the following results hold true in the very low and very high temperature limits (see \cite{SM}). 
\paragraph*{CC and magnetic field sweeps}\label{subsec:CoulombandB}-- In \eqnref{eq:phase}, $\delta R_{\text{L}} \propto \Re [e^{i \theta}]$ varies with changes in the applied magnetic field through both the $\phi$ and the $N_{\text{L}}$ terms. The dominant frequency for this variation is \cite{SM}
\vspace{-0.5mm}
\be\label{eq:abelianBsweepformula}
m = \frac{g\nu_{\text{in}}}{e_{\text{in}}^{*}}+ e_{\text{in}}^{*} - \frac{\theta_a \nu_{\text{in}}}{ 2\pi e_{\text{in}}^{*} }
\ee

in units of $A_{0} \phi_{0}^{-1}$, where $g=-\left[\left[ \frac{\kappa e_{\text{in}}^{*2}}{\Delta \nu} - \frac{\theta_a}{2\pi}\right]\!\right]$. For brevity, we drop $A_{0}$ and quote $B$-sweep frequencies in units of $\phi_{0}^{-1}$. 
%
\paragraph*{ CC and side-gate voltage sweeps}\label{subsec:CoulombandV}-- $\delta R_{\text{L}} \propto \Re [e^{i \theta}]$ varies with changes in the applied side gate voltages because this affects the area enclosed by the interfering edge (on which both $\phi$ and $N_{\text{L}}$ depend) as well as the background charge $\overline{q}$.  We employ the linearized approximation of \cite{Rosenow2011}, which assumes that the background charge varies with side-gate voltage like $\delta \overline{q} = \gamma \delta V_G$. Here $\gamma$ is a quantity we expect to be relatively constant between different plateaus. The flux in the cell is related to changes in the side-gate voltage according to $\delta \phi = (\mu/\nu_{\text{in}}) \delta V_G$. For an area gate, we expect a further relation $\gamma = (1 + \tilde{\epsilon}) \mu$ where $\tilde{\epsilon}= \frac{\overline{\nu} - \nu_{\text{in}}}{\nu_{\text{in}}}$. Eventually we will use $\tilde{\epsilon}$ as a fitting parameter which we nonetheless expect (from \cite{Rosenow2011}) to be  in the range $- \frac{\Delta \nu}{2 \nu_{\text{in}}} \leq \tilde{\epsilon}\leq  \frac{\Delta \nu'}{2 \nu_{\text{in}}}$. Here $\Delta \nu' = \nu_{\text{next}}-\nu_{\text{in}}$ and $\nu_{\text{next}}$ is the next plateau appearing in the center of the device upon decreasing the magnetic field. The dominant frequencies are
\be\label{eq:abelianVsweepformula}
 k = -\db{g - \frac{\theta_a}{2\pi}}\frac{\tilde{\epsilon}}{e_{\text{in}}^{*}} + e_{\text{in}}^{*}/\nu_{\text{in}}
\ee
in units of $\mu$, where $g$ is determined as below \eqnref{eq:abelianBsweepformula}. 

\begin{figure}[h]
\includegraphics[width=0.9\columnwidth]{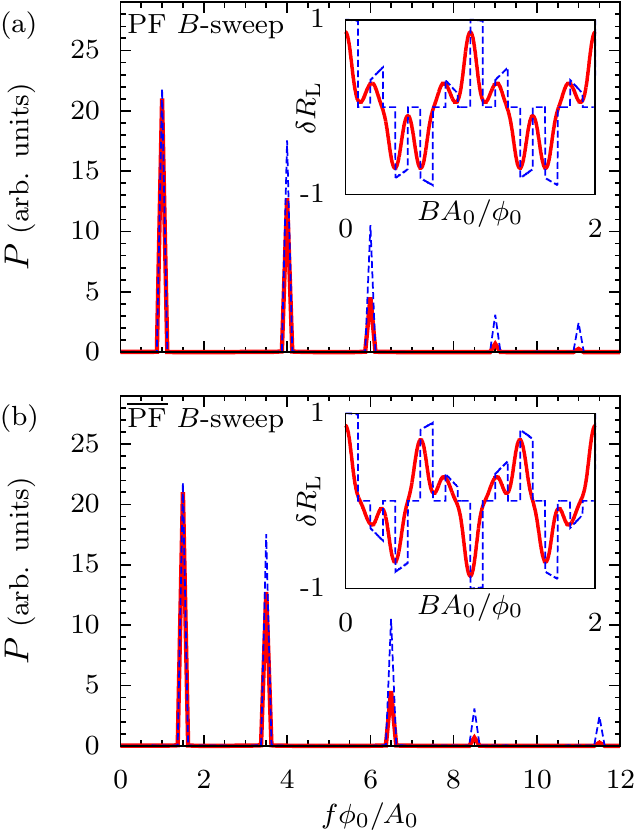}
\caption{ This figure shows the power spectrum predicted by \eqnref{eq:standard} for the variation of $\left\langle \delta R_{\text{L}}\right\rangle $ with magnetic field at $\nu=5/2$, assuming no bulk-edge Coulomb coupling and fixed fermion parity. The number of QP's was thermally averaged using the energy $E=(K/2)(N_{\text{L}} - 10 \phi/\phi_0)^2$, with a ratio of charging energy $K$ to temperature $T$ of $T/K=1/5$ for the red curve, and $T/K=0$ for the blue dotted curve. (a) and (b) label the Pfaffian and Anti-Pfaffian cases respectively. The dominant peaks at $1,1.5 \phi_{0}^{-1}$ respectively, disagree with the experimentally observed frequency $5\phi_{0}^{-1}$. The insets show the corresponding plots of $\left\langle \delta R_{\text{L}}\right\rangle $
against $B A_{0} /\phi_{0}$.}
\label{fig:Willettprediction}
\end{figure}

\paragraph*{\bf Bulk-edge Coulomb coupling at $\nu=5/2$ \label{sec:The-standard-picture}}--  Having described Coulomb effects for various abelian plateaus, we now turn our attention to $\nu=5/2$. As the physics of this plateau is less well understood, we briefly overview a number of different models for the $\nu=5/2$ device, examining these in light of the Willett {\it et al}. experimental results. We then examine the possible scenarios in the presence of CC. 

A simple model for the $\nu=5/2$ FP cell predicts a variation in longitudinal resistance \cite{Stern06,Bonderson06,Stern08}

\begin{align}
\delta R_{\text{L}} & = R_{e/4}(-1)^{N_{\psi}}\left[1+(-1)^{N_{\text{L}}}\right]\cos\left(\frac{1}{4}2\pi\frac{BA_{0}}{\phi_{0}}\pm \frac{\pi}{4}N_{\text{L}}\right)\nonumber \\
+ & R_{e/2}\cos\left(\frac{1}{2}2\pi\frac{BA_{0}}{\phi_{0}}\pm \frac{\pi}{2}N_{\text{L}}\right)\punc{,}\label{eq:standard}
\end{align}

where $N_{\text{L}}$ is defined in \eqnref{eq:NL}. Here $(-1)^{N_{\psi}}$ denotes the parity of neutral
fermion modes in the cell. The $\pm$ cases
apply to the PF/$\overline{\text{PF}}$ states respectively. 


\paragraph*{Fixed versus random $N_{\psi}$ parity regimes}-- Before investigating Coulomb effects at $\nu=5/2$, we highlight two possibilities:  $(-1)^{N_{\psi}}$ might fluctuate little on the measurement time-scale ($\sim1s$), or it may fluctuate a lot. 

 Willett {\it et al.} assume the former -- that $(-1)^{N_{\psi}}$ fluctuates little. We now argue that this is inconsistent with their data. Experiments indicate a $B$-sweep frequency of $5\phi_{0}^{-1}$, which Willett {\it et al.} attribute to the even-odd term $1+(-1)^{N_{\text{L}}}$ in \eqnref{eq:standard}. We concur that the frequency of oscillations in the $B$-sweep experimental data $5\phi_{0}^{-1}$ roughly agrees with the even-odd term $1+(-1)^{N_{\text{L}}}$. However, this term also multiplies a cosine in \eqnref{eq:standard}, and the product has a much lower dominant frequency of $1,1.5\phi_{0}^{-1}$ for the $\text{PF},\overline{\text{PF}}$ cases respectively. We illustrate this point in \figref{fig:Willettprediction}, where we also show that the dominant low frequency modes are robust at finite temperature. Neither of these dominant low frequency modes appear in the Willett \textit{et al.} interference data. Thus, the naive theoretical prediction contradicts the experimental data \cite{Willett3,Willett4,Willett5}.

Not only is fixed $(-1)^{N_{\psi}}$ inconsistent with experiment, in \cite{SM} we estimate from microscopic calculations that $(-1)^{N_{\psi}}$ should fluctuate appreciably over the measurement time scale, so we are strongly in the random $(-1)^{N_{\psi}}$ regime. In this regime, the first line of \eqnref{eq:standard} disappears on average i.e.,  there is no net $e/4$ contribution to $\delta R_{\text{L}}$ regardless of whether $N_{\text{L}}$ is even or odd. Thus we expect that any interference present comes from $e/2$
edge modes, represented by the second line in \eqnref{eq:standard}
\[
\delta R_{\text{L}} = I_{e/2} \cos\left(\frac{1}{2}2\pi\frac{BA_{0}}{\phi_{0}}\pm \frac{\pi}{2}N_{\text{L}}\right)\punc{.}
\]

Notice that the $e/2$ QP accrues only an abelian phase, which is twice the abelian phase acquired by an $e/4$ QP upon encircling the FP cell. The above function has a dominant frequency of $f=-2\phi_{0}^{-1},+3\phi_{0}^{-1}$ for the Pfaffian/Anti-Pfaffian states respectively. The latter frequency of $3\phi_{0}^{-1}$, corresponding to the Anti-Pfaffian, seems too small to be consistent with the $B$-field sweep data of Willett \textit{et al.} In summary, we have shown that both the random and fixed $N_{\psi}$ parity regimes are at odds with the experimental data. In the next section, we investigate the extent to which bulk-edge CC can reconcile theory with experiment. 

\paragraph*{Predictions for $\nu=5/2$}
\begin{figure}[h]
\includegraphics[width=.92\columnwidth]{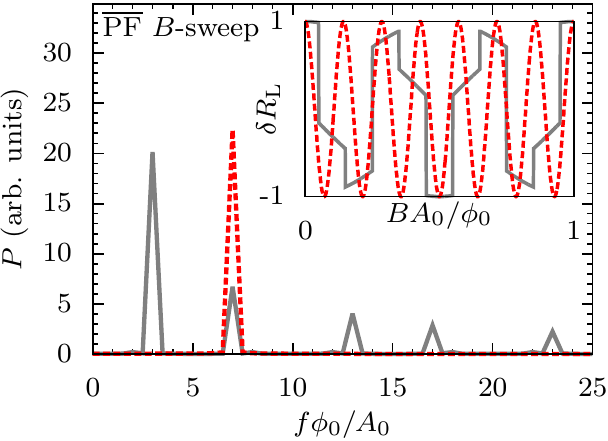}
\caption{ The inset shows the variation of $\delta R _{\text{L}}$ with magnetic field at $\nu=5/2$, assuming random fermion parity on the measurement time-scale. The main figure shows the power spectra corresponding to this variation. The grey (solid), red (dotted) curves correspond to the Aharonov-Bohm ($\kappa=0$) and Coulomb dominated ($\kappa=1$) regimes respectively. This figure shows the $\overline{\text{PF}}$ in rough agreement with experiment. See \cite{SM} for the PF case which does not fit the experimental data}\label{fig:PFCD1}
\end{figure}
-- The Coulombic correction to the $B$ and $V_{\text{G}}$ dependence of $\delta R_{\text{L}}$ for the $\nu=5/2$ plateau follows readily from \eqnref{eq:standard} and \eqnref{eq:phase}.  In this section we will consider the effect of CC on both the $\text{PF}/\overline{\text{PF}}$ states, in two regimes; one where $(-1)^{N_{\psi}}$ is random, and one where it is fixed. For clarity, we specialize to the experimentally relevant situation where $\Delta \nu =1/6$, although we provide more general expressions in \cite{SM}.

\paragraph*{$B$-sweep, Constant $N_{\psi}$}:
 For the $\text{PF}$ state, the dominant frequency is $-1\phi_{0}^{-1}$
for $0\leq\kappa<1$, but as $\kappa$ approaches to $1$ the mode $-6\phi_{0}^{-1}$ is equally prominent. In the anti-Pfaffian case, the dominant frequency is $1.5\phi_{0}^{-1}$ for $0\leq\kappa<1/3$, and $-3.5\phi_{0}^{-1}$ for $1/3<\kappa\leq1$.

\paragraph*{$B$-sweep, Random $N_{\psi}$ parity}: For the $\text{PF}$ state, the dominant frequency is $-2\phi_{0}^{-1}$
for small $\kappa$, but as $\kappa$ increases to $1$ the $-12\phi_{0}^{-1}$ mode is equally prominent. In the $\overline{\text{PF}}$ case, the dominant
frequency is $3\phi_{0}^{-1}$ for small $\kappa$, but as soon as $\kappa > 1/3$
, $-7\phi_{0}^{-1}$ is the dominant frequency (\figref{fig:PFCD1}).

We now summarize the side-gate sweep behavior similarly. In these cases, the results depend on $\tilde{\epsilon}$. For clarity, we will state the frequencies in units of the scale $\mu$.
\paragraph*{$V_{{\text{G}}}$-sweep, Constant $N_{\psi}$}:
 For the $\text{PF}$ state, the dominant frequency is $\frac{1}{10} + \frac{1}{2}\tilde{\epsilon}$
for $0\leq\kappa<1$, but as $\kappa$ approaches $1$ the mode $\frac{1}{10} + \frac{5}{2} \tilde{\epsilon}$ is equally prominent. In the $\overline{\text{PF}}$ case, the dominant frequency is $\frac{1}{10} -\frac{1}{2}\tilde{\epsilon}$ for $0\leq\kappa<1/3$, $\frac{1}{10} + \frac{3}{2}\tilde{\epsilon}$ for $1/3<\kappa\leq1$.

\paragraph*{$V_{{\text{G}}}$-sweep, Random $N_{\psi}$ parity}:
 For the $\text{PF}$ case, $\frac{1}{5}+\tilde{\epsilon} $ is the dominant frequency for small $\kappa$, but as $\kappa$ approaches $1$ there is an equally dominant $\frac{1}{5}+ 5\tilde{\epsilon}$ mode. For the $\overline{\text{PF}}$ state the dominant
frequency is $\frac{1}{5}-\tilde{\epsilon}$ for small $\kappa$, but as soon as $\kappa> 1/3$, the $\frac{1}{5}+3\tilde{\epsilon}$ mode dominates.

\paragraph*{\bf Comparison to experiments}\label{sec:comparetoexp}
-- The recent data of Willett {\it et al}. was measured in FP devices over a wide range of integer and fractional plateaus, most frequently $1 \leq \nu \leq 6$ and $\nu=7/3,5/2$. In this section we compare these data with the theoretical predictions made in the previous section. We find a good fit to experiment provided $1/3<\kappa<1/2$, with a few exceptions which we detail. We first treat the abelian plateaus and then $\nu=5/2$, focussing on both the $B$-sweep data and the $V_{\text{G}}$ data.  In the $\nu=5/2$ case we reemphasize an important point made above: the standard interpretation of the experiment as having negligible Coulomb effects and stable fermion parity is inconsistent with the $B$-sweep data. We present a model with moderate Coulomb coupling, and random fermion parity which better fits the $B$-sweep data and which is also largely consistent with the behavior at other plateaus. In the discussion that follows, we will assume that $1/3<\kappa<1/2$.

\paragraph*{$B$-sweeps in the abelian plateaus}
Willett {\it et al}. observe that for integer plateaus $1\leq \nu\leq 6$, the dominant frequency is $1\phi_0^{-1}$. Based on their data, we find it plausible that $\Delta \nu =1/3$ for $\nu=1,2,3$ and $\Delta \nu=1$ for  $\nu=4,5,6$. With these assumptions, provided $\kappa<1/2$, their data is perfectly consistent with our predictions in \eqnref{eq:abelianBsweepformula} for all plateaus except for $\nu=3$; in this special case, we predict a dominant frequency of $2\phi_0^{-1}$ rather than $1\phi_0^{-1}$. Although our prediction in this case disagrees with experiment, the experimental data is rather less compelling at $\nu=3$, and it is unclear whether the dominant frequency is  $1\phi_0^{-1} \text{ or } 2\phi_0^{-1}$ (see \cite{SM}).  The Heiblum experimental group recently reported $B$-sweep oscillations of $2 \phi_{0}^{-1}$ at $\nu=3$, which is naively consistent with the model above \cite{Heiblum14}. However,  their data (specifically the $V_{G}$ vs. $B$ contour plots) constrain $0\leq \kappa\leq 1/6$, so the $2 \phi_{0}^{-1}$ frequency seen in this experiment must have a source other than the bulk edge Coulomb coupling we consider. In the $\nu=7/3$ case, Willett {\it et al}. observe a dominant frequency of $2 \phi_{0}^{-1}$. Assuming in this case $\Delta \nu=1/3$, our predictions are consistent with their data. In summary, for the range of $\kappa$ values considered, the Coulombic model is consistent with the abelian plateaus observed in experiment \cite{Willett4,Willett5}, except in the (experimentally ambiguous) special case of $\nu=3$ discussed above. 
\paragraph*{$V_{\text{G}}$-sweeps in the abelian plateaus}
For the integer plateaus $1\leq \nu \leq 6$, as well as $\nu=7/3$, Willett \textit{et al.} claim to see dominant frequencies of approximately $\mu e_{\text{in}}^{*} /\nu$. Our predictions (assuming the same values of $\Delta \nu$) are in agreement with this observation to within $17\%$ for the integer plateaus, if we assume $\tilde{\epsilon}$ is within the expected window stated above \eqnref{eq:abelianVsweepformula}. In the $\nu=7/3$ case, our predictions are within $25\%$ of the required frequency $\mu/7$ provided $-0.035<\tilde{\epsilon}<0.04$ -- this is not so strenuous a restriction, given the theoretical restriction $-0.072<\tilde{\epsilon}<0.04$. So for moderate $\kappa$, the Coulombic model is consistent with Willett \textit{et al.}'s side-gate sweep data for the abelian plateaus with some reasonable value of $\tilde{\epsilon}$. 


\paragraph*{$B$-sweeps at $\nu=5/2$} --
 upon fixing $V_{\text{G}}$ and varying only the magnetic field, \cite{Willett3,Willett4,Willett5} observe an interference trace with dominant frequency (or a pair of dominant frequencies near) $\sim5\phi_{0}^{-1}$ \cite{Willett3,Willett4,Willett5}. In the previous section we saw that these data are {\it inconsistent} with Willett {\it et al.}'s implicit assumptions that the devices have fixed $(-1)^{N_{\psi}}$ and weak Coulomb effects -- these assumptions imply much lower dominant frequencies of $1,1.5 \phi_{0}^{-1}$ for  PF/$\overline{\text{PF}}$ respectively, which are significantly lower than the observed $\sim5\phi_{0}^{-1}$. 

The presence of two frequencies near $\sim5\phi_{0}^{-1}$ is more consistent with the $\overline{\text{PF}}$ state with random $(-1)^{N_{\psi}}$ and with moderate $\kappa$, where we showed the frequency $-7 \phi_{0}^{-1}$ dominates and $3 \phi_{0}^{-1}$ is subdominant. Thus, it seems more plausible that \cite{Willett1,Willett2,Willett3,Willett4,Willett5} observe a $\nu=5/2$ $\overline{\text{PF}}$ with random fermion parity in the bulk and moderate $\kappa$. 

\paragraph*{$V_{\text{G}}$-sweeps at $\nu=5/2$}
The experimental data power spectra show the presence of two prominent modes which have frequencies in ratio $1:2$ -- these are referred to as the `$e/4$' and `$e/2$' modes in \cite{Willett1,Willett2,Willett3,Willett4,Willett5}. The Coulomb model exhibits similar behavior for the $\overline{\text{PF}}$ state with random $N_{\psi}$ and moderate $\kappa$, although some tuning of the $\tilde{\epsilon}$ parameter is required.   If we set $\tilde{\epsilon} \in [-0.03,-0.023]$ or $\tilde{\epsilon}\in [0.029,0.05]$, we find that there are two Fourier modes present which are in ratio $2$ to within $17\%$ \footnote{ The theoretical discussion above \eqnref{eq:abelianVsweepformula} only restricts $\tilde{\epsilon} \in [-0.03,0.1]$.}.
Even with this tuning, simple Coulomb effects do not capture all aspects of the data. In particular, in experiment the dominant frequency is claimed to alternate between the two modes as the side-gate voltage is ramped. This alternation in dominant frequency could come from thermal or environmental noise, although we do not have a complete model for these effects as they appear to require a detailed understanding of the electrostatics in these devices.


\paragraph*{\bf Conclusion}\label{sec:conclusion}
-- We have detailed the effects of bulk-edge Coulomb coupling in Fabry-Perot devices, and explained how these Coulombic effects are enhanced in devices for which certain FQHE states are present. We then applied this picture to the recent experiments of Willett {\it et al}. We first noted that experiment is inconsistent with the standard picture of the even-odd effect in the FP cell -- the standard picture predicts $B$-sweep frequencies which are much lower than those observed. We then suggested a reason for why the standard picture predictions disagree with experiment:  our best estimates of the FP cells' microscopic parameters suggest that the parity of neutral fermions in the interferometer cell fluctuate significantly over the experimental time-scale, a scenario that is completely at odds with the standard picture.  We then attempted to reconcile the scenario with random fermion parity with Willett {\it et al.}'s data, and found that it is largely possible to do so if we incorporate moderate bulk-edge Coulomb coupling. Our conclusions suggest it is important to check more precisely the importance of bulk-edge Coulomb coupling in the $\nu=5/2$ devices. In particular, it would be useful to probe the crossover between the CD and AB regimes, perhaps using a range of device geometries. 
%
%

\paragraph*{\bf Acknowledgements}-- We thank Kirill Shtengel and Chetan Nayak for explaining their work to us, as well as Moty Heiblum and Michael~P.~Zaletel for useful discussions. This work was supported by the Princeton Center for Theoretical Science (CWVK), DFG grant RO 2247/8-1 (BR), and the EPSRC through a doctoral studentship (CWVK) and Grants EP/I032487/1 and EP/I031014/1 (SHS).

\bibliographystyle{apsrev4-1}
\bibliography{../thesis}

\appendix

\section{Is the bulk fermion parity random?}

In this section we discuss the stability of $(-1)^{N_{\psi}}$ in the $\nu=5/2$ PF/$\overline{\text{PF}}$ interferometer cell. There are three important frequency scales we consider. First there is
 a frequency $\omega_{\text{exp}}\sim1\text{Hz}$ associated with the experimental measurement time-scale. Second, $\omega_{\text{sd}}=\hbar e^{*}V_{\text{sd}}$ is a frequency associated with the temporal width of the edge-mode wave-packet, where $V_{\text{sd}}$ is the source-drain bias. Third and last, $\w_{\text{be}}= \mathcal{T}^{2}_{\text{be}}\times R/ v_{\text{n}}$ is the frequency of bulk edge Majorana tunneling. This incorporates the bulk edge Majorana tunneling element $\mathcal{T}_{\text{be}}$, the neutral edge mode velocity $v_{\text{n}}$ and an additional length scale $R$ on the edge which we take to be of order a few magnetic lengths; these quantities are estimated below.

It has been argued that
there is no even-odd effect for $ \w_{\text{be}}/\omega_{\text{sd}} =\frac{\mathcal{T}_{\text{be}}^{2}R}{\hbar v_{\text{n}}e^{*}V_{\text{sd}}} \gg 1$ \cite{Rosenow2008,Rosenow2009,Bishara2009,Bishara2008}.
In this regime, the $e/4$ pattern remains even when there are an
odd number of particles in the bulk, and the edge modes do not experience
any monodromy with the bulk QP's -- they acquire only an Aharonov-Bohm
phase from encircling the cell. Therefore, if the standard even-odd effect
is to arise, we need $\w_{\text{be}}/\omega_{\text{sd}} \lesssim1$.
For our best estimates of the experimental values, this ratio is actually
$\frac{R}{l_{B}}[1.056,2\times10^{-5}]$ for $\nu=5/2$ plateau region of width $w_{\text{hw}}\in[2.5,15]l_{B}$, suggesting we are in a moderate or weakly coupled regime. Thus naively one might expect to observe the even-odd effect, especially in the wider devices.

However, this does not yet imply that the even-odd effect will be
visible. For this, we require further that the bulk fermion parity fluctuates little over the course of a measurement, presumably requiring $\omega_{\text{exp}}>\omega_{\text{be}}$. However, we estimate below that $\omega_{\text{be}}/\omega_{\text{exp}} \gtrsim 10^7$, suggesting that $N_{\psi}$ is randomized over the course of an experimental measurement. Therefore the $e/4$ term in \eqnref{eq:standard} disappears
on average, regardless of whether $N_{\text{L}}$
is even or odd. We may hope that $(-1)^{N_{\psi}}$ is fixed on average thermally because the bulk qubit states have an energetic splitting. Numerics suggest \cite{Baraban09} that the energy splitting $\Delta$ between the qubit states for a pair of QP/QH's falls off approximately as $ 1 \text{K}\times e^{-r/2.3 l_B}$, suggesting that a pair of QP's/QH's
must be within a distance $r \lesssim 10l_{B}$ if their splittings are to exceed the temperature $25\text{mK}$. If there are $N$ QP's (isolated from the edge) forming a chain with regular spacing, we expect the splitting between qubits to be further suppressed by a factor of $\sim 1/N$. In any case, it can also be shown that the qubit splitting for such a chain is only as strong as its weakest odd link i.e., if an odd subsection of the chain is coupled weakly (with tunneling element $h$) to the rest of the chain, then the gap goes like $\lesssim h$.  However, we expect that the positions of the QP's in the bulk are partly determined by the random positions of donor layer ions. For this reason, we do not expect the qubit gap to reproducibly exceed $25 \text{mK}$, and so we do not expect $(-1)^{N_{\psi}}$ to be fixed thermally \footnote{In \cite{Kang11}, a different regime is discussed, but it also likely requires $\Delta \gtrsim T$.}.

In conclusion, we find it plausible that $e/4$ oscillations
should not be observed in the Willett \textit{et al.} experiments because, while the
bulk QP's could be well separated from the edge, they
are not sufficiently far from the edge to preclude the scrambling
of qubits on the experimental time-scale. 

\section{Detailed analysis of $B$ (magnetic field) sweeps}
\begin{table*}
\begin{tabular}{|c|c|c|c|c|c|c|c|c|c|c|c|c|c|}
\cline{3-14} 
\multicolumn{1}{c}{} &  & \multicolumn{2}{c|}{$0\leq\kappa<\frac{1}{6}$} & \multicolumn{2}{c|}{$\frac{1}{6}<\kappa<\frac{1}{3}$} & \multicolumn{2}{c|}{$\frac{1}{3}<\kappa<\frac{1}{2}$} & \multicolumn{2}{c|}{$\frac{1}{2}<\kappa<\frac{2}{3}$} & \multicolumn{2}{c|}{$\frac{2}{3}<\kappa<\frac{5}{6}$} & \multicolumn{2}{c|}{$\frac{5}{6}<\kappa<1$}\tabularnewline
\hline 
$\nu_{\text{in}}$ & $\Delta\nu$ & $B$ & $V_{\text{SG}}$ & $B$ & $V_{\text{SG}}$ & $B$ & $V_{\text{SG}}$ & $B$ & $V_{\text{SG}}$ & $B$ & $V_{\text{SG}}$ & $B$ & $V_{\text{SG}}$\tabularnewline
\hline 
$1$ & $\frac{1}{3}$ & $1$ & $1$ & $0$ & $1+\tilde{\epsilon}$ & $0$ & $1+\tilde{\epsilon}$ & $-1$ & $1+2\tilde{\epsilon}$ & $-1$ & $1+2\tilde{\epsilon}$ & $-2$ & $1+3\tilde{\epsilon}$\tabularnewline
\hline 
$2$ & $\frac{1}{3}$ & $1$ & $\frac{1}{2}$ & $-1$ & $\frac{1}{2}+\tilde{\epsilon}$ & $-1$ & $\frac{1}{2}+\tilde{\epsilon}$ & $-3$ & $\frac{1}{2}+2\tilde{\epsilon}$ & $-3$ & $\frac{1}{2}+2\tilde{\epsilon}$ & $-5$ & $\frac{1}{2}+3\tilde{\epsilon}$\tabularnewline
\hline 
$3$ & $\frac{1}{3}$ & $1$ & $\frac{1}{3}$ & $-2$ & $\frac{1}{3}+\tilde{\epsilon}$ & $-2$ & $\frac{1}{3}+\tilde{\epsilon}$ & $-5$ & $\frac{1}{3}+2\tilde{\epsilon}$ & $-5$ & $\frac{1}{3}+2\tilde{\epsilon}$ & $-8$ & $\frac{1}{3}+3\tilde{\epsilon}$\tabularnewline
\hline 
$4,5,6$ & $1$ & $1$ & $\frac{1}{\nu}$ & $1$ & $\frac{1}{\nu}$ & $1$ & $\frac{1}{\nu}$ & $1-\nu$ & $\frac{1}{\nu}+\tilde{\epsilon}$ & $1-\nu$ & $\frac{1}{\nu}+\tilde{\epsilon}$ & $1-\nu$ & $\frac{1}{\nu}+\tilde{\epsilon}$\tabularnewline
\hline 
$7/3$ & $\frac{1}{3}$ & $2$ & $\frac{1}{7}+\tilde{\epsilon}$ & $2$ & $\frac{1}{7}+\tilde{\epsilon}$ & $2$ & $\frac{1}{7}+\tilde{\epsilon}$ & $2$ & $\frac{1}{7}+\tilde{\epsilon}$ & $2$ & $\frac{1}{7}+\tilde{\epsilon}$ & $2$ & $\frac{1}{7}+\tilde{\epsilon}$\tabularnewline
\hline 
$5/2$ ($\overline{\mbox{PF}}$,rand) & $\frac{1}{6}$ & $3$ & $\frac{1}{5}-\tilde{\epsilon}$ & $3$ & $\frac{1}{5}-\tilde{\epsilon}$ & $-7$ & $\frac{1}{5}+3\tilde{\epsilon}$ & $-7$ & $\frac{1}{5}+3\tilde{\epsilon}$ & $-7$ & $\frac{1}{5}+3\tilde{\epsilon}$ & $-7$ & $\frac{1}{5}+3\tilde{\epsilon}$\tabularnewline
\hline 
$5/2$ (PF,rand) & $\frac{1}{6}$ & $-2$ & $\frac{1}{5}+\tilde{\epsilon}$ & $-2$ & $\frac{1}{5}+\tilde{\epsilon}$ & $-2$ & $\frac{1}{5}+\tilde{\epsilon}$ & $-2$ & $\frac{1}{5}+\tilde{\epsilon}$ & $-2$ & $\frac{1}{5}+\tilde{\epsilon}$ & $-2$ & $\frac{1}{5}+\tilde{\epsilon}$\tabularnewline
\hline 
$5/2$ ($\overline{\mbox{PF}}$,fixed) & $\frac{1}{6}$ & $1.5$ & $\frac{1}{10}-\frac{\tilde{\epsilon}}{2}$ & $1.5$ & $\frac{1}{10}-\frac{\tilde{\epsilon}}{2}$ & $-3.5$ & $\frac{1}{10}+\frac{3\tilde{\epsilon}}{2}$ & $-3.5$ & $\frac{1}{10}+\frac{3\tilde{\epsilon}}{2}$ & $-3.5$ & $\frac{1}{10}+\frac{3\tilde{\epsilon}}{2}$ & $-3.5$ & $\frac{1}{10}+\frac{3\tilde{\epsilon}}{2}$\tabularnewline
\hline 
$5/2$ (PF,fixed) & $\frac{1}{6}$ & $-1$ & $\frac{1}{10}+\frac{\tilde{\epsilon}}{2}$ & $-1$ & $\frac{1}{10}+\frac{\tilde{\epsilon}}{2}$ & $-1$ & $\frac{1}{10}+\frac{\tilde{\epsilon}}{2}$ & $-1$ & $\frac{1}{10}+\frac{\tilde{\epsilon}}{2}$ & $-1$ & $\frac{1}{10}+\frac{\tilde{\epsilon}}{2}$ & $-1$ & $\frac{1}{10}+\frac{\tilde{\epsilon}}{2}$\tabularnewline
\hline 
\end{tabular}
\caption{This table shows the $B$ and $V_{\text{G}}$ sweep (dominant) frequencies (in units of $\phi_{0}^{-1},\mu$ respectively) for a variety of plateaus $\nu_{in}$, over a range of the bulk-edge Coulomb coupling $\kappa$. In the case of $\nu=5/2$, $\text{PF}/\overline{\text{PF}}$ denote the Pfaffian/anti-Pfaffian states respectively, while `rand' and `fixed' denote the random/fixed fermion parity regimes.}\label{tab:frequencysummary}
\end{table*}%

\label{App:Bdetails}

In this section, we examine the Fourier spectra for $\delta R_{\text{L}}$ vs $B$ plots implied by \eqnref{eq:phase} for various Hall plateaus considered in \cite{Willett1,Willett2,Willett3,Willett4,Willett5}. Although the power spectra we examine are taken to be at zero temperature, the results in this section relating to which frequencies are dominant at given values of $\kappa$ hold true in the large temperature limit as well. Our results are summarized in \tabref{tab:frequencysummary}.

\subsection{Abelian plateaus}
The power spectra for general $B$-field sweeps (applying to $\nu=1,2,3,4,5,6$ and $\nu=7/3$ for instance) is
\begin{equation}
\label{eq:Bsweepgen}
p_{m}\propto\frac{\sin\left[\pi\Delta \right]^{2}}{
 \left(\Delta+ g \right)^{2}}
\end{equation}

where the frequency (in units of $\phi_{0}^{-1}$) is $m = \frac{g\nu_{\text{in}}}{e_{\text{in}}^{*}}+ e_{\text{in}}^{*} - \frac{\theta_a \nu_{\text{in}}}{ 2\pi e_{\text{in}}^{*} }$ for $g\in \mathbb{Z}$, and $\Delta=\frac{\kappa e_{\text{in}}^{*2}}{\Delta \nu} - \frac{\theta_a}{2\pi}$. The dominant frequency will correspond to the $g$ that minimizes $\left(\Delta + g \right)^{2}$. We now discuss some abelian plateaus of particular interest. 

\paragraph*{ $\nu =$  integer:}
In this case $m = 1 - \nu \dlsq \frac{\kappa }{\Delta \nu}\drsq$. For plateaus $\nu=1,2,3$, it is plausible that $\nu=2/3, 5/3, 8/3$ respectively are the  next outermost plateaus, suggesting $\Delta\nu = 1/3$. In this case we expect $\nu=1,2,3$ to have dominant frequency $1 \phi_0^{-1}$ for $\kappa < 1/6$ , and respective dominant frequencies $0,-1,-2 \phi_0^{-1}$ for   $ 1/6 < \kappa <1/2 $. For all three plateaus $\nu=4,5,6$, we expect $\Delta \nu =1$, which implies a dominant frequency $1 \phi_0^{-1}$ in all cases, for $\kappa < 1/2$.  \cite{Willett4,%
Willett5} claims to see fundamental frequency $1\phi_{0}^{{-1}}$ for all of these integer plateaus, which is mostly consistent with the above model provided  $\kappa<1/2$. The only exception is $\nu=3$, which has frequency $-2\phi_{0}^{{-1}}$ when $\kappa$ is in the interval $[1/6,1/2]$. 

\paragraph*{ $\nu =7/3$:}
Starting with \eqnref{eq:Bsweepgen},  we find $m= 5 - 7\dlsq\frac{2}{3} + \frac{\kappa}{9\Delta \nu}\drsq$. Thus, if $\kappa < 15 \Delta \nu/2$, then $-2 \phi_0^{-1}$ is the dominant frequency. As $\Delta \nu$ is likely to be $1/3$, we have $15 \Delta \nu/2 = 5/2$ and so, as $0\leq\kappa<1$, $-2 \phi_0^{-1}$ will always be the dominant frequency. Willett \textit{et al.} claim to see fundamental frequency $2\phi_{0}^{{-1}}$ for $\nu=7/3$. The Coulomb model is consistent with the observations for all values $0\leq\kappa<1$.

\subsection{$\nu=5/2$ with fixed $N_{\psi}$ parity}

In the fixed $(-1)^{N_{\psi}}$ regime, there are two kinds of modes $a=0,1$ with power spectra

\begin{equation}
\label{eq:Bsweep5o2nonrand}
p_{m_{a}}\propto \frac{\sin\left[\pi\Delta_{a}\right]^{2}}{\left(\Delta_{a}+g_{a}\right)^{2}} 
\end{equation}

for frequencies (in units of $\phi_{0}^{-1}$) of $m_{a}=10 g_{a}-\frac{5\eta}{4}+5a +\frac{1}{4}$ with $g_{a}$ integers and $\Delta_{a}= \db{\frac{\kappa}{\Delta \nu}+8 a-2\eta}/{16}$. Here $\eta=\pm 1$ for the $\text{PF},\overline{\text{PF}}$ states respectively. The dominant frequency is determined by choosing a $g_{a}$ and $a$ so as to minimize $\mid g_{a} + \Delta_{a}  \mid$. Let us assume $\Delta \nu=1/6$.  For the $\text{PF}$ state, the dominant frequency is $-1\phi_{0}^{-1}$ for $0\leq\kappa<1$, but as $\kappa$ approaches $1$ the mode $-6\phi_{0}^{-1}$ is equally prominent. In the anti-Pfaffian case, the dominant frequency is $1.5\phi_{0}^{-1}$ for $0\leq\kappa<1/3$, $-3.5\phi_{0}^{-1}$ for $1/3<\kappa\leq1$.

\subsection{$\nu=5/2$ with random $N_{\psi}$ parity}
\begin{figure}[h]
\includegraphics[width=.92\columnwidth]{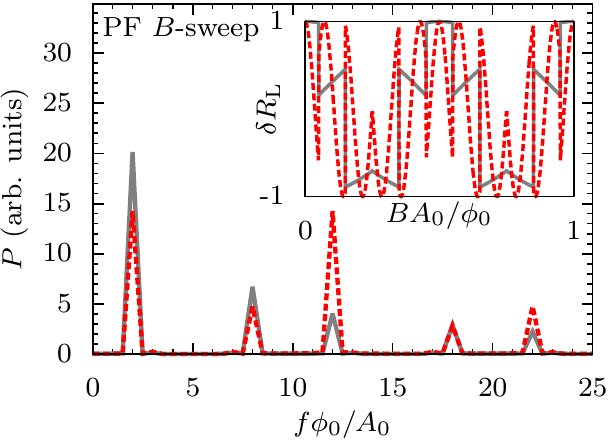}
\caption{ The inset shows the variation of $\delta R _{\text{L}}$ with magnetic field, assuming that number of neutral fermions in the cell is random over the experimental time-scale. The main figures show the power spectra corresponding to this variation. The grey (solid), red (dotted) curves correspond to the Aharonov-Bohm ($\kappa=0$) and Coulomb dominated ($\kappa=1$) regimes respectively. This figure shows the $\text{PF}$ case, see the main text for the $\overline{\text{PF}}$ which better fits the experimental data.}\label{fig:PFCD2}
\end{figure}

At the $\nu=5/2$ plateau, in a state with suppressed $e/4$ oscillations, the power spectrum is the same as in \eqnref{eq:Bsweepgen}
%
except $m=10 g+\frac{1}{2}-\frac{5\eta}{2}$ for $g\in \mathbb{Z}$, and $\Delta=\db{\frac{\kappa}{\Delta \nu}-2\eta}/8$. In what follows, we will assume that $\Delta \nu=1/6$, in accordance with experiment. In the Pfaffian case ($\eta=1$), the dominant frequency is $-2\phi_{0}^{-1}$
for small $\kappa$, but as $\kappa$ increases to $1$ the $-12\phi_{0}^{-1}$
mode is just as prominent (\figref{fig:PFCD2}). In the Anti-Pfaffian case ($\eta=-1$), the dominant
frequency is $3\phi_{0}^{-1}$ for small $\kappa$, but for $1/3 < \kappa < 1$
, $-7\phi_{0}^{-1}$ is the dominant frequency (see \figref{fig:PFCD1}).


\section{A detailed analysis of $V_{\text{G}}$ (side gate voltage) sweeps}\label{App:Vdetails}
\begin{figure}[htbp]
\begin{center}
\includegraphics[width=.95\columnwidth]{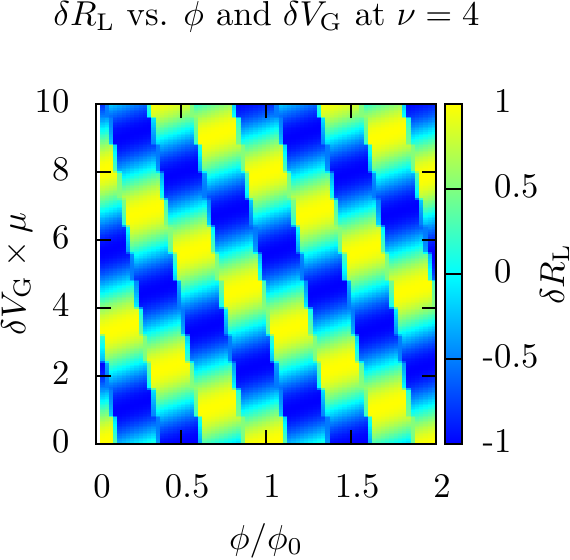}
\caption{This figure shows a contour plot for $\nu=4$ at $\kappa=1/3$ with $\Delta \nu=1$, with the color scale denoting variations in longitudinal resistance $\delta R_{\text{L}}$ and the axes corresponding to flux through the cell $\phi/\phi_{0}$ and side-gate voltage $\delta V_{\text{G}}$ (arb. units). The negative slope of the diagram is consistent with that found in Fig.~6 of \cite{Willett4}.}
\label{fig:nu4contourplot}
\end{center}
\end{figure}

In this section, we examine the power spectra for the $\delta R_{\text{L}}$ vs $V_{\text{G}}$ plots implied by \eqnref{eq:phase}. We consider various Hall plateaus considered in \cite{Willett1,Willett2,Willett3,Willett4,Willett5}. We briefly recap some details of the side-gate sweeps. Changing the side gate voltage $V_{\text{G}}$ changes both the
area of the cell $A$ and the level of background charge $\bar{q}$, and therefore leads to a change in both
$\phi$ and $\bar{q}$. We remark here that $\overline{q}$ changes for two reasons: i) the charge density inside the layer changes (which we neglect later on), and ii) the area of the interior changes, giving rise to a change in the background charge as well. Following the
prescription in \cite{Rosenow2011}, we will assume that $\delta A=\frac{\beta\phi_{0}}{B}\delta V_{\text{G}}$. Notice that $\beta = \frac{1}{\nu_{\text{in}}} \mu$, where we expect that $\mu$ is relatively constant between different plateaus. For an area gate we find $\delta \bar{q}=\mu (1+\tilde{\epsilon}) \delta V_{\text{G}}$. For our comparison to experiments we use $\tilde{\epsilon}$ as a fitting parameter which we nonetheless expect (from \cite{Rosenow2011}) to be  in the range 
\be \label{eq:boundsonepsilon}
- \frac{\Delta \nu}{2 \nu_{\text{in}}} \leq \tilde{\epsilon}\leq  \frac{\Delta \nu'}{2 \nu_{\text{in}}}\punc{.}
\ee
 Our results are summarized in \tabref{tab:frequencysummary}.

\subsection{Abelian plateaus}
The Fourier modes for $\nu=1,2,3,4,5,6$ and $\nu=7/3$ have frequencies $k = -(g - \frac{\theta_a}{2\pi})\frac{\tilde{\epsilon}}{e_{\text{in}}^{*}} + e_{\text{in}}^{*}/\nu_{\text{in}}$ in units of $\mu$, where $g\in \mathbb{Z}$. The power spectra for these modes is the same function of $g$ as \eqnref{eq:Bsweepgen}. The dominant frequency will, as for the $B$-field sweeps, correspond to the $g$ that minimizes $\left(\frac{\kappa e_{\text{in}}^{*2}}{\Delta \nu} - \frac{\theta_a}{2\pi} + g \right)^{2}$. 

\paragraph*{ $\nu =$  integer:}
In this case $k = \frac{1}{\nu_{\text{in}}} + \tilde{\epsilon} \dlsq\frac{\kappa }{\Delta \nu}\drsq$. Using \eqnref{eq:boundsonepsilon}, the dominant frequency lies in the range
\be
 \frac{1}{\nu_{\text{in}}} - \frac{\Delta \nu}{2 \nu_{\text{in}}}\dlsq\frac{\kappa }{\Delta \nu}\drsq \leq k \leq   \frac{1}{\nu_{\text{in}}} +\frac{\Delta \nu'}{2 \nu_{\text{in}}}  \dlsq\frac{\kappa }{\Delta \nu}\drsq\punc{.}
\ee

The predictions are as follows. $\nu=1,2,3$ have dominant frequencies $\frac{1}{\nu}   \mu$  respectively for $\kappa < \Delta \nu/2=1/6 $ , and roughly the same frequency for  $ 1/6 < \kappa <1/2 $, although the precise value of this frequency will depend on $\tilde{\epsilon}$ -- Assuming that $\Delta \nu' = \Delta \nu$ for each of these plateaus, we find that the frequency is between $\frac{\mu}{\nu_{in}} (1\pm \frac{1}{6})$. 

For the plateaus $\nu=4,5,6$, we expect $\Delta \nu =1$  so that the plateaus have frequencies $\frac{1}{\nu}  \mu$ for $\kappa < \Delta \nu/2=1/2$ regardless of the value of $\tilde{\epsilon}_{\nu}$. Thus, we expect that provided $ \kappa <1/2 $, the ratio of side gate frequencies will be largely the same as in the AB ($\kappa=0$) limit. As an aside, note that  $\nu=4$ is still in its AB regime for $1/3<\kappa<1/2$ and so this range of $\kappa$ values is consistent with the negative slope shown in the plot (compare \cite{Willett4} fig 6 with \figref{fig:nu4contourplot}). 


\paragraph*{ $\nu =7/3$:}
Here we find that $k= \frac{1}{7}+(-2+3 \dlsq \frac{2}{3}+ \frac{\kappa}{9 \Delta \nu}\drsq)\tilde{\epsilon}$. This reduces to $k= \frac{1}{7}+\tilde{\epsilon}$ assuming $0\leq \kappa\leq1$ and $\Delta \nu = 1/3$. If we use $\Delta \nu=1/3$ and $\Delta\nu'=1/6$ with  \eqnref{eq:boundsonepsilon}, then the dominant frequency is between $\frac{\mu}{7}[1-\frac{1}{2},1+\frac{1}{4}]$ for all $0<\kappa<1$ .

\subsection{$\nu=5/2$ }

In this section, we will find that the dominant frequencies depend quite sensitively on the $\tilde{\epsilon}$ parameter. Using \eqnref{eq:boundsonepsilon} with $\Delta \nu =1/6$ and $\Delta \nu'=1/2$ we get
\be\label{eq:epsilonrange5/2}
- \frac{1}{30} \leq \tilde{\epsilon}\leq  \frac{1}{10}\punc{.}
\ee

\subsubsection{Fixed $N_{\psi}$ parity}

As in \eqnref{eq:Bsweep5o2nonrand}, there are two types of Fourier mode $a=0,1$. These  have frequencies $k_{a}=-(4g+2a-\frac{\eta}{2})\tilde{\epsilon}+\frac{1}{10}$ in units of $\mu$, for integer $g_{a}$. The power spectra are the same functions of $g_{a}$ as \eqnref{eq:Bsweep5o2nonrand}, so the dominant frequencies are once more determined by the $g_{a}$ values closest to $\Delta_{a}= \db{\frac{\kappa}{\Delta \nu}+8 a-2\eta}/{16}$.

For the $\text{PF}$ state, the dominant frequency is $\frac{1}{10} + \frac{1}{2} \tilde{\epsilon}$
for all $0\leq\kappa<1$ -- using \eqnref{eq:epsilonrange5/2}, this frequency lies in range $\frac{\mu}{10}[1-\frac{1}{6},1+\frac{1}{2}]$. In the $\overline{\text{PF}}$ case, the dominant frequency is $\frac{1}{10} -\frac{1}{2}\tilde{\epsilon}$ for all $0\leq\kappa<1/3$, which lies in the range $\frac{\mu}{10}[1-\frac{1}{2},1+\frac{1}{6}]$ if we use \eqnref{eq:epsilonrange5/2}. The dominant frequency is $\frac{1}{10} +\frac{3}{2} \tilde{\epsilon} $ for all $1/3<\kappa\leq1$, which similarly lies in the range $\frac{\mu}{10} [1-\frac{1}{2},1+\frac{3}{2}]$.

\subsubsection{Random $N_{\psi}$ parity}

The frequency (in units of $\mu$) is now $k=-(4g-\eta)\tilde{\epsilon}+\frac{1}{5}$. The dominant frequencies are once again determined by the integers nearest $\Delta=\db{\frac{\kappa}{\Delta \nu}-2\eta}/8$. Again using \eqnref{eq:epsilonrange5/2}, we list the dominant frequencies. In the Pfaffian case, the dominant frequency is $k=\frac{1}{5}+\tilde{\epsilon}$ for $0\leq\kappa<1$, which is in $\frac{\mu}{5}[1-\frac{1}{6},1+\frac{1}{2}]$. In the Anti-Pfaffian case, the dominant frequency is $k=\frac{1}{5}-\tilde{\epsilon}$ for $0\leq\kappa < 1/3$ which lives in range $\frac{\mu}{5}[1-\frac{1}{2},1+\frac{1}{6}]$. For $1/3 < \kappa <1$ it is $\frac{1}{5}+3\tilde{\epsilon}$,  which lives in $\frac{\mu}{5}[1-\frac{1}{2},1+\frac{3}{2}]$.

\section{High temperatures limits}\label{app:largeT}
It is straightforward to generalize the work of \cite{Rosenow2011} to large temperatures i.e., $T\geq K_{\text{I}}, K_{\text{L}}, K_{\text{IL}}$. As was found in that work, the question of which frequency dominates is actually the same in both the high and low temperature limits where one can analytically find the Fourier transforms.

\subsubsection{Abelian plateaus}
In the large temperature limit, we agree with the results of \cite{Rosenow2011} that the power spectra for both the $B,V_{\text{G}}$ sweeps take form

\be\label{eq:highT}
\exp\ds{-2\pi^{2}T \db{ \frac{e_{\text{in}}^{2} \alpha^{2} }{K_{{\text{I}}}^{2} \Delta \nu^{2}  }  + \frac{\db{g+ \Delta }^{2} K_{\text{I}}}{e_{\text{in}}^{2} (K_{\text{I}}K_{\text{L}}-K_{\text{IL}}^{2})}    }}
\ee

where the $B$-sweep frequency $m$ and the $V_{{\text{G}}}$-sweep frequencies $k$ are related to $g$ in the same way as in the zero temperature case. We set $\alpha=1$ for abelian plateaus. We see that for a given value of the bulk edge Coulomb coupling $\kappa$, the dominant frequency is determined by the $g$ value closest to $\Delta$. Hence, the dominant frequencies for both $B,V_{\text{G}}$ depend on $\kappa$ precisely as they did in the zero temperature limit. 

\subsubsection{$\nu=5/2$ fixed $N_{\psi}$ parity}

The same pattern holds true for the $\nu=5/2$ plateaus. In the fixed $(-1)^{N_{{\psi}}}$ regime there are again two kinds of modes $a=1,2$ with power spectra\be\label{eq:highTa}
\exp\ds{-2\pi^{2}T \db{ \frac{e_{\text{in}}^{2} }{K_{{\text{I}}}^{2} \Delta \nu^{2}  }  + \frac{\db{g_{a}+ \Delta_{a} }^{2} K_{\text{I}}}{e_{\text{in}}^{2} (K_{\text{I}}K_{\text{L}}-K_{\text{IL}}^{2})}    }}
\ee
where $g_{0},g_{1}$ are integers and $\Delta_{a}= \frac{\frac{\kappa}{\Delta \nu}+8 a-2\eta}{16}$. The $B,V_{\text{G}}$ frequencies have the same dependence on $g_{a}$ as at zero temperature, and the dominant frequencies are again determined by choosing $g_{a}$ and $a$ so as to minimize $\mid g_{a}+\Delta_{a}\mid$. This gives the same dominant frequencies as in the zero temperature cases.

\subsubsection{$\nu=5/2$ random $N_{\psi}$ parity}
In this case the power spectra take the form shown in \eqnref{eq:highT} except one sets $\alpha=2$, and $\Delta=\db{\frac{\kappa}{\Delta \nu}-2\eta}/8$. Again, the dominant frequency depends on $\kappa$ in the same way as in the zero temperature regime.

\section{Estimates}\label{app:estimates}

\begin{table}
\begin{tabular}{|c|c|}
\hline 
$e^{*}V_{\text{sd}}$ & $1.5\times10^{-2}\text{K}$\tabularnewline
\hline 
$\mathcal{T}_{\text{be}}$ & $[0.3,60]\text{GHz}$\tabularnewline
\hline 
$\w_{\text{exp}}$ & $1\text{Hz}$\tabularnewline
\hline 
$\frac{\mathcal{T}_{\text{be}}^{2}R}{\hbar v_{\text{n}}e^{*}V_{\text{sd}}}$ & $\frac{R}{l_{B}}[1.1,2.1\times10^{-5}]$\tabularnewline
\hline 
\end{tabular}
\caption{This table summarizes estimates of the important scales in the Hall device around the $\nu=5/2$ plateau.}
\label{tab:quantities}
\end{table}

In this section we estimate the values of several important physical parameters in the Willett {\it et al.} devices. The results are summarized in \tabref{tab:quantities}.

The active area of the FP cells fall in the range $[0.1,0.6] \mu \text{m}^2$ \cite{Willett5}. The lengths of the devices, on the other hand, are around $2\mu \text{m}$. This suggests that the half-width $w_{\text{hw}}$ of the $\nu=5/2$ plateau is in range $[0.026,0.15]\mu \text{m}$ or approximately $[2.5,15] l_B$. If we assume the $\nu=5/2$ regime is described by a $\text{PF}/\overline{\text{PF}}$ state, then numerics \cite{Rezayi09} indicate a neutral edge mode velocity of $v_{\text{n}}\sim 10^4 \text{ms}^{-1}$. 

The source drain bias $V_{\text{sd}}$ is obtained directly from \cite{Willett1,Willett2,Willett3,Willett4}, where it is stated that the current through the device is of the order $2 \text{nA}$. Each current carrying quantum hall edge has a resistance of $\nu h/e^2$, so we estimate the device resistance by estimating the net effects of the edge modes present at the $\nu=5/2$ plateau $R = \frac{h}{e^2} (5/2 + 7/3 + 2 +\ldots  )^{-1}$, which gives $e^{*}V_{\text{sd}} \approx 15 \text{mK}$, associated with frequency-scale $1.9 \text{GHz}$. 

The bulk-edge hopping element $\mathcal{T}_{\text{be}}$ is estimated using results from Monte-Carlo calculations \cite{Baraban09}, which suggests a qubit splitting of form $1.76\times10^{11}\exp(-r/2.3 l_{B})\text{GHz}$. In the absence of better estimates, we assume the bulk-edge splitting takes the same form, and we set $r$ to be the distance from bulk QP's/QH's to the edge. Note $r$ is bounded above by $w_{\text{hw}}$ (the half-width of the device), so we take $r$ to lie in the range $[2.5,15] l_B$. Using this range, we estimate that $\mathcal{T}_{\text{be}}$ is in $[0.3,60]\text{GHz}$.

The last quantity to estimate is the effective bulk-edge Majorana tunneling rate $\frac{\mathcal{T}_{\text{be}}^{2}R}{\hbar v_{\text{n}}e^{*}V_{\text{sd}}}$. This quantity was derived in \cite{Rosenow2009} from a lattice regularized Majorana chain model, where $R$ was interpreted as the lattice spacing. In the Hall device there is no equivalent lattice spacing scale, so we assume that $R\sim l_B$ which is the only natural scale in the system. 

\end{document}